\documentclass[a4paper,11pt]{article}
\usepackage[utf8]{inputenc}
\usepackage[british]{babel}
\usepackage[mono=false]{libertine}
\usepackage[T1]{fontenc}
\usepackage[top=2.5cm, bottom=2.5cm, left=2cm, right=2cm]{geometry}
\usepackage[font=small]{caption}
\usepackage{subcaption}
\usepackage{cite}
\usepackage{verbatim}
\usepackage{enumitem}
\usepackage{titlesec}
\usepackage[usenames, dvipsnames]{color}
\usepackage[titletoc]{appendix}
\usepackage[pdftex]{graphicx}
\usepackage{array}
\usepackage{url}
\usepackage{mathtools}
\usepackage{amssymb,amsfonts,amsmath,amsthm}
\usepackage{dsfont}
\usepackage{bm}
\usepackage[perpage,hang,flushmargin]{footmisc} 
\usepackage{enumitem}
\usepackage{booktabs}       
\usepackage{nicefrac}       
\usepackage{microtype}      
\usepackage{tikz}
\usepackage[ruled,vlined]{algorithm2e}
\usepackage{wrapfig}
\usepackage{xr,xr-hyper}
\usepackage{cases}
\usepackage{setspace}
\usepackage[pdfpagemode=UseNone,bookmarksopen=false,colorlinks=true,urlcolor=blue,citecolor=magenta,citebordercolor=blue,linkcolor=blue]{hyperref}
\usepackage{cleveref}
\makeatletter
\newcommand{\setappendix}{Appendix~\thesection:~~}
\newcommand{\setsection}{\thesection~~}
\titleformat{\section}{\bfseries\LARGE}{%
	\ifnum\pdfstrcmp{\@currenvir}{appendices}=0
	\setappendix
	\else
	\setsection
\fi}{0em}{}
\makeatother
\DeclareUnicodeCharacter{00A0}{~}

\DeclareMathOperator{\Tr}{Tr}

\theoremstyle{definition}

\makeatletter
\renewcommand\paragraph{\@startsection{paragraph}{4}{\z@}%
                                      {\parskip}
                                      {-1em}%
                                      {\normalfont\normalsize\bfseries}}
\makeatother

\newcommand{\cs}{\mathcal{S}}
\newcommand{\vcs}{\vec{\mathcal{S}}}
\newcommand{\cF}{\mathcal{F}}
\newcommand{\vcF}{\vec{\mathcal{F}}}
\newcommand{\cN}{\mathcal{N}}
\newcommand{\vx}{\vec x}
\newcommand{\vy}{\vec y}
\newcommand{\va}{\vec a}

\newcommand{\veta}{\vec\eta}
\newcommand{\vm}{\vec m}
\newcommand{\vn}{\vec n}
\newcommand{\gb}[1]{\textcolor{red}{#1}}

\allowdisplaybreaks

\begin{document}
\setcounter{tocdepth}{2}
\setcounter{secnumdepth}{3}
\title
{Generative diffusion in very large dimensions}
\maketitle

\begin{center}
  Giulio Biroli \textsuperscript{1} and Marc Mézard \textsuperscript{2}
\end{center}
\begin{center}
$^1$ Laboratoire de Physique de l'Ecole Normale Sup\'erieure, ENS, Universit\'e PSL, CNRS, Sorbonne Universit\'e, Universit\'e Paris-Diderot, Sorbonne Paris Cit\'e, Paris, France\\ 
$^2$ Department of Computing Sciences, Bocconi University
\end{center}

\maketitle
\date{}
\begin{abstract}
Generative models based on diffusion have become the state of the art in the last few years, notably for image generation. 
Here, we analyse them in the high-dimensional limit, where data are formed by a very large number of variables. We use methods from statistical physics and focus on two well-controlled high-dimensional cases: a Gaussian model and the Curie-Weiss model of
ferromagnetism. In the latter case, we highlight the mechanism of symmetry breaking in the inverse diffusion, and point out that, in order to reconstruct the relative asymmetry of the two low-temperature
states, and thus to obtain the correct probability weights, one
needs a database with a number of points much larger than the dimension of each data point. We characterize the scaling laws in the number of data and in the number of dimensions for an efficient generation.  
\end{abstract}


%
\renewcommand{\labelitemi}{$\bullet$}

In recent years, machine learning has made spectacular progress in building powerful and efficient generative models. In particular, diffusion models have emerged as powerful tools for modeling complex data distributions and generating realistic samples. They have become the state of the art in generating images, videos or sound
{\cite{Sohl_Dickstein2015,Song2019,Song_Sohl-Dickstein2021,guth2022wavelet,yang2022diffusion}}. 
These models leverage diffusion processes to transform simple, tractable initial distributions into complex target distributions. While the empirical success of generative diffusion models has been widely acknowledged, a comprehensive theoretical understanding of their underlying principles remains elusive.  

The underlying process is guaranteed to be successful for finite dimensional data and general bounds assessing their performances have been obtained \cite{de2021diffusion,lee2022convergence,de2022convergence}. However,  an analysis explicitly taking into account the high-dimensionality of realistic data is lacking. This is a crucial issue since  
subtle questions arise when generating probability distributions in very large dimensional spaces. One of the most striking examples is the concentration of the probability measure in very different portions of the configuration space, which in physics correspond to symmetry-breaking and phase transitions. 

In this work we address this challenge and 
investigate the performance and the limitations of generative diffusion models when they are used for data living in a $N\gg 1$ dimensional space, with a diffusion potential estimated from $P$ data points.  In particular, we want to understand (i) scaling laws: the typical range of $P$ that is needed when $N$ is large, and (ii) selective perfomance: the ability of diffusion models to reproduce some but not all aspects of high-dimensional probability laws depending on the scaling of $P$ with $N$.
We explore these issues in two models: a
model of high-dimensional gaussian data where the covariance matrix has a well-defined
density of eigenvalues, and a probabilistic model of binary variables (Ising spins) with long-range interactions, the
Curie-Weiss model of a ferromagnet, which has a phase transition at low temperature. \\

Let us start by describing the general setup for generative
diffusion. We study $N$-dimensional variables, $a=\{a_1,...,a_N\}$, distributed
according to a probability law $P_0(a)$. We suppose to be given
$P$ iid samples
of $a$: $a^1,...,a^P$, where $a^\mu=\{a_1^\mu,...,a_N^\mu\}$, called the
data. 
Generative diffusion aims at generating new samples of $a$ efficiently,
sampled from a very accurate estimation of $P_0$. It can be decomposed into two phases, a training phase and a generative phase.

During {\it training}, one generates
$N$-dimensional trajectories, starting from each of the available
data points, which are evolved in time  through independent
Ornstein-Uhlenbeck processes. This is a ``forward'' diffusion
process, in which, for each $\mu$ we
have $\vx^\mu(t=0)=\va^\mu$, and then each $\vx$ evolves in $\mathbb{R}^N$ according to the
Langevin equation
\begin{align}
\frac{d\vx}{dt}=-\vx+\veta(t)
\end{align}
where $\veta(t)\in\mathbb{R}^N$ is a Gaussian process with zero mean and $\langle
\eta_i(t)\eta_j(t')\rangle=2 T \delta_{ij} \delta(t-t')$.

The parameter $T$ controls the amount of noise in the Ornstein-Uhlenbeck process. We keep here
to the simplest version in which the confining potential is constant
in time (here taken as $\vx^2/2$) and the variance of the noise is also
time-independent. More general diffusion processes where the confining
potential and the noise variance
 also evolve in time are often used for practical
applications and could be studied as well \gb{\cite{yang2022diffusion}}. One could also generate several trajectories starting from
each data point. 

After time $t$, the forward process generates points distributed
according to the probability $P_t(\vx)$ given by
\begin{align}
P_t(\vx)=\int d\va \; P_0(\va) \frac{1}{\sqrt{2\pi
  \Delta_t}^N}\exp\left(-\frac{1}{2} \frac{(\vx-\va
  e^{-t})^2}{\Delta_t}\right) = \int d\va \; P_t(\va,\vx)
\end{align}
which is the convolution of the original distribution with a Gaussian law of variance
$\Delta_t=T(1-e^{-2t})$. We have also defined here  $ P_t(\va,\vx)$, the joint distribution of
$\va$ and $\vx(t)$ for a given $t$.

A crucial quantity for the generating process
is the gradient of the log of the probability to be in any point
in the $N$ dimensional space, which is called the score \cite{Song2019}. 
Fixing the time $t$, the score is a vector field with components
\begin{align}
\cF_i(\vx,t)=\frac{\partial \log P_t(\vx)}{\partial x_i}= -
  \frac{x_i-\langle a _i\rangle_{\{x\}}e^{-t}}{\Delta_t}
  \label{exact-score}
\end{align}
where $ \langle a _i\rangle_{\{x\}}$  is the average of $a_i$ with
respect to the conditional probability $P_t(\va|\vx)=P_t(\va,\vx)/P_t(\vx)$.
In practice, in most cases one does not really know this score exactly,
and one approximates it by a score function $\vcs^{\theta (t)}(\vx) $ which is a
$N$-dimensional vector field, parametrized through a neural network architecture defined by some parameters
$\theta (t)$,  which are estimated from the $P$ data points . The parameters $\theta(t)$ are obtained by minimizing a loss
function 
which measures the difference between $\vcs^{\theta }(\vx) $
and $\vcF(x,t)$. For instance one can take a square loss:
\begin{align}
\mathcal{L}(\theta)= \int d\vx P_t(\vx)
  \left\lVert\vcs^{\theta}(\vx) -\vcF(\vx,t)\right\rVert^2\,\, .
  \label{squareloss}
\end{align}
This can be rewritten as
\begin{align}
  \mathcal{L}(\theta)={\mathbb{E}}_{x,a}\left\lVert\vcs^\theta(\vx)+\frac{\vx-\va
  e^{-t}}{\Delta_t}\right\rVert^2+C
  \end{align}
where $ {\mathbb{E}}_{x,a}$ denotes the expectation with respect to
the joint probability $P_t(\vx,\va)$, and the constant $C$ is
independent of $\vcs^\theta$. This is a convenient form since it can be
estimated empirically from the database points $\va^\mu$ and their evolved position $\vx^\mu$ \cite{Song2019}. In realistic applications all losses for different times $t$ are packed together with suitable weights,  to define a global loss {\cite{ho2020denoising} (the weights are chosen in order to improve generation quality and simplicity of implementation)}. This setting could be studied as well. 
At a large enough final time $t_f$ (when the  Ornstein-Uhlenbeck process, which relaxes exponentially fast, has reached equilibrium), the endpoints of these trajectories are thermalized in the
quadratic potential, so they become iid
variables with Gaussian distribution $G_T(\vx)=e^{-\vx^2/(2T)}/\sqrt{2\pi
  T}^N$.

Let us now describe the second phase of the process, the {\it generation} phase. One first generates points according to the 
distribution, $G_T(\vx)$ and let them evolve through an inverse diffusion that takes time backward, from $t_f$ to
$0$ (we consider $t_f$ large enough so that for all practical purposes $\vx$ has thermalized) . It is a mathematical fact\cite{Anderson1982reverse-diff} that, if this backward Langevin process
is driven by a force 
which is the sum of the exact score plus the inverse of the confining force used in the forward process, then the backward diffusion
will generate at $t=0$ points distributed according to
$P_0$. Precisely, if we start at $t=t_f$ from $N$ variables $y_i$ which are
iid Gaussian distributed according to $\cN(0,T)$, and let them evolve
through the backward Langevin process defined by
\begin{align}\label{eq:tr}
-\frac{dy_i}{dt}=y_i+2 T \cF_i(y,t) +\eta_i(t)
\end{align}
where the noise $\eta(t)$ has the same distribution as in the forward
process, then the vector $\vy(0)$ is distributed according to
$P_0$. Actually at each time $t$, the points $\vy$ generated by the
backward process have the same distribution $P_t$ as the points $\vx$
generated by the forward process: the probability law is transported back in time by the process (\ref{eq:tr}). This can be shown using the Fokker-Planck equation.

With respect to this ideal mathematical setup, the performance of generative
diffusion in large dimensional space ($N\gg 1$) can be limited by
several factors, the main ones being
\begin{enumerate}
\item
  The interpolating model used to find the approximate score function
  $\vcs^{\theta}(\vx)$ might not be able to represent the exact 
  score  $\cF(\vx,t)$ accurately enough.
  \item
  In order to estimate the score, we have access to trajectories
  generated by a finite number of initial points $P$. This finite
  sampling of the probability $P_0$ will induce some
  imprecision in the score.
  \end{enumerate}

A simple example of the first point is when one uses a linear
regression in order to estimate the score. Let us study this case,
assuming that we have infinite data, so that we have access to the
full forward probability. For a
given time $t$ of the forward process, we 
search a score function in the form
\begin{align}
\cs_i(x, t)=-\sum_j W_{ij} x_j-b_i
  \end{align}
where the parameters $W$ and $b$ are determined to minimize the 
loss. Using the square loss (\ref{squareloss}), one finds that the
optimal values of $W$ and $b$ are given by
  \begin{align}
     W&= \left(T(1-e^{-2t}){\mathbb{I}}+e^{-2t}C^0\right)^{-1}\\
    b_i&=-e^{-t}\sum_j W_{ij} m_j^0
         \label{Wopt}
  \end{align}
  where $\vm^0$ and $C^0_{ij}$ are the mean and covariance of the
  initial distribution $P_0$. Because the score function is linear and
  the  backward evolution starts at $t_f$ from a Gaussian
  distribution, it  constructs at time $t=0$ a Gaussian
  distribution $\tilde P$. If the parameters $W$ and $b$ have been chosen
  optimally as in (\ref{Wopt}), $\tilde P$ is a Gaussian measure which
  has the same mean $\vm$ and covariance $C^0$ as the distribution
  $P_0$ that we started from.  
  But this in general is not enough: such a simple linear score cannot  reconstruct non-Gaussian distributions.  In realistic applications very large neural networks are used to represent very general classes of score function {\cite{yang2022diffusion}}. 
  
  We now turn to the study of the second point, which is the main focus of this work, and concerns
  undersampling : how large should the size $P$ of the database
  be? We start with a detailed analysis of a high-dimensional Gaussian model.
We assume that the distribution $P_0$ is Gaussian with mean zero and covariance
$C^0$. We call $K^0=(C^0)^{-1}$.
  Then at each time the measure $P_t$ of the forward process $\vx$ is Gaussian,
with mean zero and covariance $C^t=(K^t)^{-1}$, where
\begin{align}
K^t= \frac{K^0}{e^{-2t}+\Delta_t K^0}\,\, .
\end{align}

Now let us consider the sampling errors. Empirically we have access to
$P$ initial points $\va^\mu$ and the corresponding evolved points at
time $t$, called $\vx^\mu(t)$. The score is computed from empirical correlation matrices (see SI), in particular using the empirical 
covariance $D^t$:
\begin{align}
D^t_{ij}&= \frac{1}{P}\sum_\mu x_i^\mu(t) x_j^\mu(t)\,\,.
  \end{align}
  Let us first study the errors due to finite sampling at time
  $t=0$. These depend a lot on the covariance of the probability $P_0$.  
When $N\gg 1$, a frequent situation is that the eigenvalues of $C^0$ are distributed with a density
such that the typical spacing between them is $1/N$. What value of $P$
is needed in order that $D^0$ be a good approximation of $C^0$ ? Let
us assume that $P$ scales as $P=N^{a+1}$.
 In order to get the same eigenvalues and
 eigenvectors for $C^0$ and $D^0$ one needs to have $a>1$, while $a>0$
 is enough to get correctly the eigenvalue-dependent observables,
 such as the density of eigenvalues or normalized traces of powers of
 the matrix. A simple argument that allows to understand this scaling goes as follows:
 It has been shown \cite{facoetti2016non,benigni2020eigenvectors,von2019non}
 that if one considers the
 sum of a given $N\times N$ matrix $A$ (with eigenvalues of the order
 of one) and a
 random matrix $G$ taken from a Gaussian Orthogonal Ensemble (with eigenvalues of the order of one),
$
H=A+\frac{1}{N^{a/2}} G
$,
then the density of eigenvalues of $H$ converges to the one of $A$ in
the large $N$ limit for $a>0$.  In order to get the same eigenvalues
and eigenvectors for $H$ and $A$ one needs to have $a>1$. These results can be understood using perturbation theory, and showing that  $\frac{1}{N^{a/2}} G$ is a negligible perturbation for the trace of the resolvent for $a>0$, whereas it is a negligible perturbation for the whole spectral properties in the case $a>1$. For $0<a<1$ the density of eigenvalues is correct but eigenvectors hybridize if they are at a distance of the order of $1/N^a$. 
The same phenomenon takes place for  $D^0$ compared to $C^0$. In fact, for each single term of $D^0$ one can write: 
\begin{equation}
[D^0]_{ij}=[C^0]_{ij}+[\delta C^0]_{ij}
\end{equation}
where the last term represents the fluctuations around the mean is of the order $1/\sqrt{P}$.  Therefore we can write
$
D^0=C^0+\sqrt{\frac{N}{P}}R=C^0+ \frac{1}{N^{a/2}} R
$
where $R$ is a random matrix with eigenvalues of the order of one
which thus matches the known scaling (see SI for a more thorough argument). One can actually repeat this argument for all correlations matrices that determine the score, and hence also for the covariance matrix of the generative process. In consequence, one expects that the diffusion model generates multivariate Gaussian samples whose covariance coincides with the original one only for 
$P\gg N^2$ (see SI for a more detailed argument). For $N\ll P \ll N^2$, observables which only depend on the eigenvalue (the strength of the fluctuations) of the covariance are correct, e.g. normalized traces of the correlation matrix, whereas the ones which depend on the eigenvectors (the directions of the fluctuations) are not. Interestingly, these two transitions also appear when evaluating different distances between the learned distribution and the true distribution\footnote{We thank an anonymous referee for suggesting this idea.}.  The 2-Wasserstein distance, suitably normalized, goes to zero when $P\gg N$ whereas the total variation distance vanishes for $P\gg N^2$ (see the Appendix for some details).

We now turn to another subtle point of probabilities in large
dimensions, namely phase transitions. If $P_0(\va)$ is in a low temperature phase, with several pure states,
how can generative diffusion
reproduce the shattering of the large dimensional space of configuration into these lumps? How can it reproduce spontaneous symmetry breaking? We will study here the
simple example of the Curie Weiss model, written as 
   \begin{align}
P_0(a)=\frac{1}{Z}\exp\left(\frac{\beta}{2N}(\sum_i
     a_i)^2+\frac{h}{N}\sum_i a_i\right)
   \end{align}
  where $a_i=\pm 1$  are $N$ Ising spins, at equilibrium  at inverse temperature $\beta$, with a
  small external field $h/N$ that creates an asymmetry. For $\beta>1$,
  this model has a phase transition. We shall see that the
  generative diffusion is able to reproduce it, but also that special care
  is needed in order to generate the exact degree of asymmetry, i.e. the correct fraction of the configurations in the positively and negatively magnetized states. This entails that the generative model can be in regimes in which typical configurations are generated correctly but their weights are not.
  
   The distribution $P_0(a)$ can be written by introducing the
   auxiliary variable $m$ through a Hubbard-Stratonovitch transfomation:
 \begin{align}
P_0(\va)=\int dm P_0(m,\va)\ \ ; \ \ P_0(m,\va) =\frac{1}{Z}  e^{-\beta N m^2/2} \prod_i e^{(\beta m+h/N)
   a_i}\,\,.
 \end{align}
 Summing over $a_i$, we can deduce the distribution of $m$,
 $P_0(m)=\sum_{\va} P_0(m,\va) \simeq Ce^{-\beta N\Phi(m)}$, where the large deviation function is 
 \begin{align}
     \Phi(m)=m^2/2-\frac{1}{\beta} \log \cosh(\beta m +h/N).
 \end{align} In the large $N$ limit, $P_0(m)$ is the sum
 of two gaussians centered in $m=\pm m^*$, with  variances of order
 $1/N$, where $m^*$ satisfies the mean-field equation
$
m^*=\tanh(\beta m^*)
 $.
 We shall consider the case of a Curie Weiss model in its
 low-temperature phase, $\beta>1$, so that $m^*>0$ and we have two
 pure states of magnetization $\pm m^*$. Actually, to simplify the
 analysis, it is convenient to consider the following simplified
 Curie Weiss model, in which the distribution of $m$ is the sum of two $\delta$ functions, instead of the sum over two gaussians with width that go to zero at large $N$. This modified model has the same large $N$ limit. The model is defined by
 \begin{align}
   P_0(m)= W_+ \delta(m-m^*)+W_-\delta(m+m^*)\ \ ; \ \
   W_\pm=\frac{e^{\pm h m^*}}{2\cosh( hm^*)}\ \ ; \ \
   P_0(\va|m)=\frac{1}{Z}\prod_i e^{\beta m a_i}
 \end{align}

 Let us first study the exact score, as defined in (\ref{exact-score}).
 The joint distribution of $\vx(t)$ and $\va$ is:
\begin{align}
  P_t(\vx,\va)=W_+P_t^+(\vx,\va)+W_-P_t^-(\vx,\va)
\end{align}
where 
\begin{align}
  P_t^\pm(\vx,\va)=\prod_j\left[
  \frac{e^{-(x_j^2+e^{-2t})/(2\Delta_t)}}{\sqrt{2\pi\Delta_t}}\right]
  \prod_j\left[
  \frac{e^{a_j (\pm\beta m^*+x_j e^{-t}/\Delta_t)}}{2\cosh(\beta m^*)}
  \right]\,\,.
\end{align}
{Note that by integrating out the $xs$ one finds that the marginal distribution of the $as$ is equal to $P_0(a)$ (by construction).}
From this expression, one can deduce the conditional expectation value of $\va$
given $\vx$ :
\begin{align}
 \langle a _i\rangle_{\{x\}}=\frac{W_+ Q_+(x) \tanh\left( \beta m^*+  x_i\frac{e^{-t}}{\Delta_t}\right)+W_-Q_-(x) \tanh\left( -\beta m^*+  x_i\frac{e^{-t}}{\Delta_t}\right) }{W_+ Q_+(x) +W_-Q_-(x) }
 \label{ai_mean}
\end{align}
where
\begin{align}
  Q_\pm(x)=  \prod_j\left[
  1\pm m^*\tanh\left( x_j\frac{e^{-t}}{\Delta_t}\right)
  \right]\,\,.
\end{align}
This expression simplifies in the regime of large times, when for all
$x_j$ one has $x_j e^{-t}/\Delta_t\ll 1$. We shall see below that the phase transition is generated in the backward process in this large-time regime, where one can linearize the $\tanh$ in order to write
\begin{align}
  Q_\pm(x)= \exp\left[\sum_j\log\left(
  1\pm m^*\tanh\left( x_j\frac{e^{-t}}{\Delta_t}\right)\right)
  \right] 
  \simeq
   \exp\left[\pm m^* \sum_j x_j\frac{e^{-t}}{\Delta_t}
  \right] =
e^{\pm N m^* M(x) e^{-t}/\Delta_t}
\label{Qpm_simple}
\end{align}
with
$
M(x)=\frac{1}{N}\sum_i x_i
$.

To leading order, we can also substitute, in the numerator of (\ref{ai_mean}), 
\begin{align}
\tanh(\pm \beta m^*+  x_ie^{-t}/{\Delta_t})\simeq
\tanh(\pm \beta m^*)=\pm m^*\ .
\label{tanh_simple}
\end{align}
Using the simplifications (\ref{Qpm_simple},\ref{tanh_simple}) in the expression (\ref{ai_mean}) of the conditional mean of $a$, we deduce the exact score function (\ref{exact-score}) at large times:
\begin{align}
S_E(x)_i=-\frac{x_i}{\Delta_t}+ m^*\frac{e^{-t}}{\Delta_t}\tanh\left[m^*\left( h+ N M(x)
  \frac{e^{-t}}{\Delta_t}\right)\right]
  \label{score_CW_simple}\,\,.
\end{align}

We can now consider the backward equation and sum the process over all
the directions. 
We introduce $\mu(t)=(1/\sqrt{N})\sum_i x_i$, where the $1/\sqrt{N}$ provides the natural scaling of the fluctuations in the short time evolution of the backward process, where the 
symmetry breaking takes place. This variable $\mu(t)$
satisfies the backward Langevin equation
  \begin{align}
-\frac{d \mu}{dt}=-\mu+2m^* \sqrt{N} e^{-t} \;
  \tanh\left[m^*\left(
  h+\frac{\sqrt{N}}{T} e^{-t} \mu\right)
  \right]+ \eta(t)
  \label{Langevin_mu}=-\frac{dV}{d\mu}+\eta(t)
  \end{align}
  where $V(\mu,t)$ is a time dependent potential, which, in the large
  time limit $t\gg 1$, is equal to
  \begin{align}
V=\frac{1}{2}\mu^2-2 T\log \cosh \left[m^*\left(h+\frac{\sqrt{N}}{T}e^{-t}\mu\right)\right]\,\,.
  \end{align}
  
It is interesting to notice that this is the  backward
diffusion potential,  for a generative
diffusion process of a single variable $\mu$ which
would be generated at time $0$ from a distribution
$
  P_0(\mu)= W_+ \delta(\mu-\sqrt{N} m^*)+W_- \delta(\mu+\sqrt{N} m^*)
$.
Therefore we can study how the two ferromagnetic phases are reconstructed by the
generative process, through an analysis  of this simple one-dimensional
stochastic process for $\mu$, in the large-time limit.
The potential $V(\mu,t)$ changes its shape, from a double-well structure to a single-well structure, at times of order $(1/2) \log N$. In fact, at 'short' times,  when $\sqrt{N} e^{-t}\gg 1$, the potential goes to 
\begin{equation}
    V_s(\mu)=\frac{1}{2}\mu^2-2 m^*\sqrt{N} e^{-t} |\mu|\ ,
\end{equation} a symmetric double well structure with a large barrier  of size $N e^{-2t}$ at the origin. On the other hand, at 'long' times,  when $\sqrt{N} e^{-t}\ll 1$, the potential goes to 
\begin{equation}
V_s(\mu)=\frac{1}{2}\left[\mu- m^*\tanh(h m^*) \frac{\sqrt{N} e^{-t}}{T}\right]^2+\text{constant}\ ,
\end{equation}
which is a quadratic well shifted from the origin at a distance proportional to the asymmetry $\tanh (h m^*)$. Therefore the generation of the phase transition in the backward diffusion process takes place at time $t\sim (1/2) \log N$. Decreasing the time, i.e. going backward in time, the barrier between the two wells forms, and becomes larger and larger compared to the stochastic noise. 

To study this transition, one can
compute the probability $R_+(t)$ that $\mu>0$ in a
regime of time
$
t=\frac{1}{2}\log N +\tau\ .
$
One finds
\begin{align}
R_+(\tau)=\frac{1}{2}\left[W_+\text{erfc}\left(-\frac{m^*
  e^{-\tau}}{\sqrt{2T}}\right)+ W_-\text{erfc}\left(\frac{m^*
  e^{-\tau}}{\sqrt{2T}}\right)\right]
  \end{align}
  {where $\text{erfc}$ is the complementary error function:
  \begin{align}
\text{erfc}=\frac{2}{\sqrt{\pi}} \int_x^\infty du e^{-u^2}\,\,.
  \end{align}}
Clearly, if $\tau\to +\infty$, we get $R_+(\tau)\to 1/2$. On the other
hand, for $\tau\to -\infty$, we get $R_+(\tau)\to e^{m^*h}/(2
\cosh(m^*h))$. This confirms that, when $N$ is very large, the transition from
the asymmetric to the symmetric distribution happens when $\tau $ is
around 0, i.e. $e^{-t}$ is of order $1/\sqrt{N}$. On this timescale, we see that the non-trivial part of the score
function for component $x_i$, the one that drives the asymmetry and is responsible for creating the correct weights of the two states,
is of order $1/\sqrt{N}$. Therefore we need a precision in the
computation of the score that is better than this. As the uncertainty
on the score is $1/\sqrt{P}$, one needs $P\gg N$ to reconstruct the
asymmetry properly. 

We have thus found  that the generative diffusion is able to reconstruct typical configurations belonging to the two states of the Curie Weiss model as soon as $P\gg 1$, provided the forward process is followed up to times $t$ such that $\sqrt{N}e^{-t}\ll 1$. However, in order to reconstruct the relative weights of these two states, one needs a much larger data base, with a number of points $P\gg N$.  

In this note, in order to have a fully solvable system we have focused on the Curie-Weiss model, but we expect this phenomenon to be generically present when the high-dimensional probability law is concentrated on a finite number of lumps, at least for generic statistical mechanics models.  In fact, imagine an initial distribution $P_0(\va)=(1/Z_0) e^{-\beta H_0(\va)}$ which is decomposed in a finite number of pure states where, in each pure state $\alpha$, the magnetizations are $\langle a_i\rangle_\alpha=m_i^\alpha$, and the free energy is $F_\alpha$. In this case, the partition function can be written as:
\[
Z=\sum_\alpha e^{-\beta F_\alpha}
\]
so that the weight of pure state $\alpha$ is $w_\alpha=e^{-\beta F_\alpha}/\sum_\gamma e^{-\beta F_\gamma} $. We consider, as previously, a situation in which the weights differs by a quantity of order one. The score can be generically related to the derivative of the free-energy $F$ of the system with respect to an external magnetic field 
$z_i=\frac{x_i e^{-t}}{\beta\Delta_t}$, where the magnetic field changes the energy $H_0(\va)$ to $H_0(\va)-\sum_iz_i a_i$. Using (\ref{exact-score}) one finds 
\begin{align}
S_E(x)_i=-\frac{x_i}{\Delta_t}- \frac{e^{-t}}{\Delta_t}\left.\frac{\partial F}{\partial z_i}\right|_{z_i=\frac{x_ie^{-t}}{\beta\Delta_t}}\,\,.
  \label{score_general_simple}
\end{align}

At large times $z_i$ is very small since $\frac{x_i e^{-t}}{\beta\Delta_t}\ll 1$, therefore one has:
\[
e^{-\beta F}=\sum_\alpha e^{-\beta F_\alpha+\beta\sum_i z_i m^{\alpha}_i}\,\,\, .
\] 
All the $F_\alpha$s are equal to leading order in $N$, but  they differ of quantities of order one, and so do the relative weights $w_\alpha=e^{-\beta F_\alpha}/\sum_\gamma e^{-\beta F_\gamma} $.  Therefore one finds that the score can be written as 
\begin{align}
S_E(x)_i=-\frac{x_i}{\Delta_t}+ \frac{\partial}{\partial x_i}\ln \left( 
\sum_\alpha w_\alpha e^{ N M_\alpha(x) \frac{e^{-t}}{\Delta_t}
}
\right)
  \label{score_general_simple2}
\end{align}
where $M_\alpha(x)=\frac{1}{N} \sum_i x_i m^\alpha_i$.  This is a generalization of eq. (\ref{score_CW_simple}) which is valid for many pure states and for general statistical mechanics models\footnote{We focus on models in which the original variables $a_i$ are binary; our approach and results can be extended to continuous variables.}.  

We can use in the present case the same analysis of the backward dynamics that we did above in the Curie-Weiss model, starting from eq. (\ref{score_CW_simple}): at the beginning of the backward process the configuration $\vec x$ is almost orthogonal to the magnetization $m_\alpha$, i.e. it has a scalar product of order $1/\sqrt{N}$.  The dynamical regime of the backward dynamics when the symmetry breaking takes place leads to the same scaling laws identified for the Curie Weiss model, thus hinting at a large degree of universality of our result. 

We expect the phenomena we have analyzed in this paper to be present also in practical applications.  
A thorough analysis by numerical experiments  on realistic cases is left for a future work.

\section*{Acknowledgement}

We thank J.P. Bouchaud, D. Chafai and V. De Bortoli for discussions. GB acknowledges
support from the ANR PRAIRIE. MM acknowledges financial support by the PNRR-PE-AI FAIR
project funded by the NextGeneration EU program.

\section{Supplementary Information}
In the following we give more details on the analysis of the case of multivariate gaussians in large dimensions. 

\subsection{Empirical Score in the multivariate Gaussian case}
In the case studied in the main text,  where the data are drawn from a multivariate Gaussian distribution with zero mean,  the score function is linear. For a
given $t$ one searches a score function in the form
\begin{align}
\cs_i(x, t)=-\sum_j W^{(t)}_{ij} x_j
  \end{align}
where the parameters $W$ are determined to minimize the empirical
loss.
The loss is then
\begin{align}
  \mathcal{L}^{e}(W)= \frac{1}{P}\sum_{\mu=1}^P
  \sum_{i=1}^N\left[(-W x^\mu)_i+ \frac{x^\mu_i-a^\mu_i e^{-t}}{T(1-e^{-2t})}\right]^2+\text{constant}
\end{align}
and can be rewritten as
\begin{align}
\mathcal{L}^{e}(W) =\Tr (W D^{t} W^T)-\frac{2}{T(1-e^{-2t})} \left[\Tr
  (D^{t} W)-e^{-t}\Tr( M^{t} W)\right]+\text{constant}\,\,.
\end{align}
Where $D^{t}$ is the empirical correlation at time $t$, and $M^{t}$ is
the empirical memory of initial condition at time $t$, defined by
\begin{align}
  D^{t}_{ij} &=\frac{1}{P}\sum_\mu x_i^\mu(t) x_j^\mu(t)\\
   M^{t}_{ij} &=\frac{1}{2P}\sum_\mu \left(x_i^\mu(t) x_j^\mu(0)+ x_i^\mu(0) x_j^\mu(t)\right)\,\,.
\end{align}

Then the optimal value of $W$ obtained by linear regression is given
in terms of these empirical correlation and memory matrices by:
\begin{align}
W=\frac{1}{T(1-e^{-2 t})} \left(\mathbb{I}-e^{-t} ( D^{t})^{-1}
  M^{t}\right)
  \label{Wlin}\,\,.
\end{align}
Notice that, when $P\to \infty$, $D^{t}\to e^{-2t}C^0+T(1-e^{-2t})$ and $M^t\to e^{-t} C^0$, and we get back the formula for the exact linear score. 
\subsection{P vs N scalings for the multivariate Gaussian case}
Here we give more detailed arguments for the two scaling regimes discussed in the text.  These arguments are not fully rigorous,  as they rely on some (very reasonable) assumptions. We think that they can be turned in a full proof following the guidelines below; we leave this challenge for a future work. \\
Let us first focus on the $P\gg N^2$ regime.  We proceed as explained in the main text by decomposing the matrices $M^t$ and $D^t$, used to compute the empirical score,  in an average and a fluctuation, e.g. for $D^t$:
$$
D^t=C^t+\sqrt{\frac{N}{P}}R^t=C^t+ \frac{1}{N^{a/2}} R^t\,\,.
$$
It is natural to assume that the matrix $R^t$, has eigenvalues of order of one and eigenvectors which are delocalized in the $C^t$-basis, thus impliying that the matrix $R^t$ has elements scaling with $N$ as in the GOE case (as it can be readily checked in simple cases, e.g. the Wishart which corresponds to $t=0$). 
One can then use standard perturbation theory to first and second order to study the eigenvalues and eigenvectors of $D^t$ and $M^t$.   
For $P\gg N^2$ one finds that eigenvalues are perturbed less than their mean-level spacing (of order $1/N$) and eigenvectors are changed in a sub-leading way.  In consequence, in this regime the empirical 
fluctuations are indeed just small perturbations, and the 
score matrix $W^{(t)}$ coincides with the exact one up to a sub-leading contribution.  

Let us now consider the regime  $N\ll P\ll N^2$. In this case $R^t$ is a relevant perturbation for the eigenvalues and eigenvectors: the former are perturbed much more than the mean-level spacing causing a scrambling of the eigenvectors which are no more oriented in the same direction of the ones obatined for $P\to \infty$.  However,  we expect that the diffusion model reproduces well the density of eigenvalues. The reason is that all the matrices involved in the score, hence the score itself, and the correlation matrix of the end of the backward process can be written as their exact value plus a remainder  $\frac{1}{N^{a/2}} R$.  The matrices $R$ are again assumed to have eigenvalues of order of one and eigenvectors which are delocalized (in the basis diagonalizing the exact matrix obtained for $P\to \infty$).  The density of eigenvalues of a matrix $C$ can be obtained in the large $N$ limit from the trace of the resolvent matrix :
\[
G(z)=\frac{1}{N}Tr \frac{1}{z-C}\,\,.
\] 
By plugging in this expression the correlation matrix at the end of the backward process, and writing it as the 
exact one plus a remainder $\frac{1}{N^{a/2}} R$, one finds that the latter produces correction to $G(z)$ which are sub-leading for $P\gg N$: they are at most of  order $N^{a/2}$ whereas $G(z)$ is of order one.  Hence, for $P\gg N$,  the trace of the resolvent of the empirical correlation matrix at the end of the backward process coincides with its exact ($P\to \infty$) counterpart, leading to the same density of eigenvalues. 

As we stressed before,  our arguments rely on some (reasonable) assumptions.  In the case of the empirical correlation matrix ($t=0$), one can check that these assumption indeed hold by generalizing the approach of \cite{facoetti2016non} to Wishart matrices.   

\subsection{Distances between empirical and true distribution in the Gaussian case}
Here we give a few details on the analysis of the distances between empirical and true distribution in the Gaussian case. 

We start focusing on the 2-Wasserstein distance, $W_2$, used in optimal transport \cite{panaretos2019statistical}. We evaluate the distance between the true Gaussian distribution and the Gaussian one associated to the generative process\footnote{The generated distribution, conditioned on the forward process, is Gaussian since the score is linear.}. 
Their respective covariances will be denoted $C$ and $D$.
Since both distributions are Gaussian, the distance can be written in terms of $C$ and $D$ \cite{givens1984class}:
\[
W_2=\frac{1}{N} \text{Tr} [D+C-2(C^{1/2}DC^{1/2})^2]\,\,,
\]
where we have introduced a factor $1/N$ to have a distance normalized to one in the large dimensional limit. 
Recalling the result of the previous sections, $D$ can be written as $C+R/N^{a/2}$. The matrix $M=C^{1/2}DC^{1/2}$ therefore reads: 
\[
M=C^2+\frac{C^{1/2}RC^{1/2}}{N^{a/2}}\,\,.
\]
Using the same argument of the previous sections, one can conclude that the density of eigenvalues of $M$ is the same of $C^2$ for $a>0$. In this case, all leading factors in $W_2$ cancel out and $W_2\rightarrow 0$ as $N\rightarrow 0$. 

Let us now consider the total variation distance $d_{TV}$. Following the result of \cite{devroye2018total}, the total variation distance between the true Gaussian distribution and the Gaussian one associated to the generative process satisfies:
\[
\frac{1}{100}\text{Min}\{1,\sum_i^N \lambda_i^2 \}\le d_{TV}\le\frac{3}{2}\text{Min}\{1,\sum_i^N \lambda_i^2 \}
\]
where $\lambda_i$ are the eigenvalues of the matrix $D\,C^{-1}-\mathbf{I}$. This matrix is equal to $R\,C^{-1}/N^{a/2}$ and its eigenvalues are of order $N^{-a/2}$. For $a>1$, one finds $\sum_i^N \lambda_i^2 \sim O(N^{1-a})\rightarrow 0$. Therefore the total variation distance, which is generically bounded between  zero and one, indeed goes to zero for $P>>N^2$.

\bibliographystyle{plain}
\bibliography{DM.bib}

\begin{thebibliography}{10}

\bibitem{Anderson1982reverse-diff}
Brian~DO Anderson.
\newblock Reverse-time diffusion equation models.
\newblock {\em Stochastic Processes and their Applications}, 12(3):313--326,
  1982.

\bibitem{benigni2020eigenvectors}
Lucas Benigni.
\newblock Eigenvectors distribution and quantum unique ergodicity for deformed
  wigner matrices, 2020.

\bibitem{de2022convergence}
Valentin De~Bortoli.
\newblock Convergence of denoising diffusion models under the manifold
  hypothesis.
\newblock {\em arXiv preprint arXiv:2208.05314}, 2022.

\bibitem{de2021diffusion}
Valentin De~Bortoli, James Thornton, Jeremy Heng, and Arnaud Doucet.
\newblock Diffusion schr{\"o}dinger bridge with applications to score-based
  generative modeling.
\newblock {\em Advances in Neural Information Processing Systems},
  34:17695--17709, 2021.

\bibitem{devroye2018total}
Luc Devroye, Abbas Mehrabian, and Tommy Reddad.
\newblock The total variation distance between high-dimensional gaussians with
  the same mean.
\newblock {\em arXiv preprint arXiv:1810.08693}, 2018.

\bibitem{facoetti2016non}
Davide Facoetti, Pierpaolo Vivo, and Giulio Biroli.
\newblock From non-ergodic eigenvectors to local resolvent statistics and back:
  A random matrix perspective.
\newblock {\em Europhysics Letters}, 115(4):47003, 2016.

\bibitem{givens1984class}
Clark~R Givens and Rae~Michael Shortt.
\newblock A class of wasserstein metrics for probability distributions.
\newblock {\em Michigan Mathematical Journal}, 31(2):231--240, 1984.

\bibitem{guth2022wavelet}
Florentin Guth, Simon Coste, Valentin~De Bortoli, and Stephane Mallat.
\newblock Wavelet score-based generative modeling, 2022.

\bibitem{ho2020denoising}
Jonathan Ho, Ajay Jain, and Pieter Abbeel.
\newblock Denoising diffusion probabilistic models.
\newblock {\em Advances in neural information processing systems},
  33:6840--6851, 2020.

\bibitem{lee2022convergence}
Holden Lee, Jianfeng Lu, and Yixin Tan.
\newblock Convergence for score-based generative modeling with polynomial
  complexity.
\newblock {\em arXiv preprint arXiv:2206.06227}, 2022.

\bibitem{panaretos2019statistical}
Victor~M Panaretos and Yoav Zemel.
\newblock Statistical aspects of wasserstein distances.
\newblock {\em Annual review of statistics and its application}, 6:405--431,
  2019.

\bibitem{Sohl_Dickstein2015}
Jascha Sohl-Dickstein, Eric Weiss, Niru Maheswaranathan, and Surya Ganguli.
\newblock Deep unsupervised learning using nonequilibrium thermodynamics.
\newblock In {\em International Conference on Machine Learning}, 2015.

\bibitem{Song2019}
Yang Song and Stefano Ermon.
\newblock Generative modeling by estimating gradients of the data distribution.
\newblock {\em Advances in Neural Information Processing Systems}, 2019.

\bibitem{Song_Sohl-Dickstein2021}
Yang Song, Jascha Sohl-Dickstein, Diederik~P Kingma, Abhishek Kumar, Stefano
  Ermon, and Ben Poole.
\newblock Score-based generative modeling through stochastic differential
  equations.
\newblock In {\em International Conference on Learning Representations}, 2021.

\bibitem{von2019non}
Per von Soosten and Simone Warzel.
\newblock Non-ergodic delocalization in the rosenzweig--porter model.
\newblock {\em Letters in Mathematical Physics}, 109:905--922, 2019.

\bibitem{yang2022diffusion}
Ling Yang, Zhilong Zhang, Yang Song, Shenda Hong, Runsheng Xu, Yue Zhao,
  Yingxia Shao, Wentao Zhang, Bin Cui, and Ming-Hsuan Yang.
\newblock Diffusion models: A comprehensive survey of methods and applications.
\newblock {\em arXiv preprint arXiv:2209.00796}, 2022.

\end{thebibliography}

  \end{document}